\documentclass[twocolumn,twoside,slac]{revtex4}
\usepackage{graphicx}
\usepackage{fancyhdr}
\usepackage{amsmath}

\newcommand{\bc}{\begin{center}}
\newcommand{\ec}{\end{center}}
\newcommand{\be}{\begin{equation}}
\newcommand{\ee}{\end{equation}}
\newcommand{\bea}{\begin{eqnarray}}
\newcommand{\eea}{\end{eqnarray}}
\newcommand{\bi}{\begin{itemize}}
\newcommand{\ei}{\end{itemize}}
\newcommand{\bd}[1]{\begin{dinglist}{#1}}
\newcommand{\ed}{\end{dinglist}}
\newcommand{\bt}{\begin{tabular}}
\newcommand{\et}{\end{tabular}}
\def\ape{\textsf{APE}}
\def\acento{\textsf{APE100}}
\def\amille{\textsf{APEmille}}
\def\anext{\textsf{apeNEXT}}

\begin{document}


\title{The apeNEXT project}


%[irisa]
\author{F.~Bodin}
\affiliation{IRISA/INRIA, Campus Universit\'{e} de Beaulieu, Rennes, France},

%[orsay]
\author{Ph.~Boucaud, J.~Micheli, O.~Pene}
\affiliation{LPT, University of Paris Sud, Orsay, France}

%[roma1]
\author{N.~Cabibbo, F.~Di~Carlo, A.~Lonardo, S.~de~Luca, 
        F.~Rapuano, D.~Rossetti, P.~Vicini}
\affiliation{INFN, Sezione di Roma, Italy}

%[parma]
\author{R.~De Pietri, F.~Di~Renzo}
\affiliation{Physics Department, University of Parma and
             INFN, Gruppo Collegato di Parma, Italy}

%Desy ztn
\author{H.~Kaldass, N.~Paschedag, H.~Simma}
\affiliation{DESY Zeuthen, Germany}

\author{V.~Morenas}
\affiliation{LPC, Universit\'{e} Blaise Pascal and IN2P3,
             Clermont, France}

%[NIC]
\author{D.~Pleiter}
\affiliation{NIC/DESY Zeuthen, Germany}

%[pisa]
\author{L.~Sartori, F.~Schifano, R.~Tripiccione}
\affiliation{Physics Department, University of Ferrara and INFN, Sezione di Ferrara, Italy}

\begin{abstract}
We present the current status of the \anext{} project. Aim of this project
is the development of the next generation of \ape{} machines which will
provide multi-teraflop computing power. Like previous machines, \anext{} is
based on a custom designed processor, which is specifically optimized
for simulating QCD. We discuss the machine design, report
on benchmarks, and give an overview on the status of the software
development.
\end{abstract}

\maketitle

%%%%%%%%%%%%%%%%%%%%%%%%%%%%%%%%%%%%%%%%%%%%%%%%%%%%%%%%%%%
%%
%%
%%%%%%%%%%%%%%%%%%%%%%%%%%%%%%%%%%%%%%%%%%%%%%%%%%%%%%%%%%%

\thispagestyle{fancy}


\section{INTRODUCTION}

\ape{} is one of several projects (for a review see \cite{1}) in the
theoretical physics community that have developed massively
parallel, high-performance computer architectures. The driving
force why physicists develop and build computers by themselves is the
success of numerical simulations in understanding the interactions of
elementary particles, in particular their strong interactions
described by quantum chromodynamics (QCD).  In the absence of
closed-form analytical solutions for theories, like QCD, one of the
most interesting approximation schemes is a reformulation of the
theory on a discrete lattice (see \cite{2} for a short introduction,
or \cite{3}). The original theory is recovered as the lattice spacing
$a$ goes to zero. This approach, pioneered by K. Wilson more than 25
years ago \cite{4}, is the starting point for Lattice Gauge Theory
(LGT). This discrete and computer-friendly formulation of quantum
field theory has triggered an immense activity.
%%see%% (See \cite{5} for an overview.)   

The phenomena investigated with such simulations range
from the permanent confinement of quarks inside hadrons to the
cosmological phase transition that occurred in the early phases of the
universe or in matter under extreme conditions as produced in
heavy-ion collision experiments. Within the framework of LGT,
fundamental parameters of QCD, like the masses of quarks or the
strength of the running strong coupling constants, have been computed
from first principles. Also, theoretical concepts such as spontaneous
chiral symmetry breaking and even the mathematical structure of the
theory itself can be tested with modern simulation techniques.  One of
the big challenges is the determination of weak matrix elements of
hadronic states to understand the interplay between weak and strong
interactions.  Problems like the $\Delta I=1/2$ rule or the violation
of CP symmetry are still open. The study of the heavy quark
semileptonic decays is crucial for the determination of the
Cabibbo-Kobayashi¨-Maskawa angles which are basic parameters of the
Standard Model. A further example is the non-leptonic decay
$K\rightarrow \pi\pi$, relevant to understand CP violation
\cite{7}. 

Many of the current LGT projects focus on the simulation of QCD with
dynamical fermions. Because of the limits in available computing power
one is forced in many cases to apply the so-called quenched
approximation, where the effects of vacuum fermion loops are
neglected. Although the currently available computing resources allow
to relieve this approximation, it will be extremely hard to lower the
masses of the dynamical quarks towards their physical values. It will
be even more difficult to reduce the lattice spacing and to do
simulations closer to the continuum limit.  A tremendous amount of
computer power is required to overcome these limitations. A panel of
the European Committee for Future Accelerator (ECFA), which proposed
an ambitious research program for the coming years, estimates that
European research groups would need well over 10 TFlops of compute power
\cite{8}.

In order to make these computing resources available at a
reasonable price, various research groups have engaged in the
development of supercomputers which are specifically optimized for
their applications.  In this paper we describe the Array Processor
Experiment (\ape{}) project, which was started in the mid eighties by
the Istituto Nazionale di Fisica Nucleare (INFN) and is now carried
out within the framework of a European collaboration with DESY and the
University of Paris Sud.

The structure of this paper is as follows: in the next section we
briefly cover the older members of the \ape{} family. We then describe in some detail
\amille{}, the \ape{} generation currently used in physics production simulations.
Subsequently, we discuss
the architecture of \anext{}, the new generation of \ape{} systems.
This is the most important part of our paper, followed by  
a short discussion of the \anext{} software environment.
The paper ends with some concluding remarks.

%% ----------------------------------------------------------------
%% ----------------------------------------------------------------
%%
%%
%%
%% ----------------------------------------------------------------
%% ----------------------------------------------------------------

\section{THE FAMILY OF \ape{} MACHINE}

\begin{table*}[t]
\caption{Key parameter comparison of the \ape family of supercomputer.
\label{tab:apefamily}}
\bt{|l|c|c|c|c|}
\hline
\hline & ~~\ape\ (1988)~~ & ~~\acento (1993)~~& ~~\amille\ (1999)~~ & ~~\anext\ (2003-)~~ \\
\hline
\hline Architecture     & SIMD  & SIMD & SIMD  & SPMD \\
\hline \# nodes         & 16 & 2048  &  2048  &  4096  \\
\hline Topology         & flexible 1D   & rigid 3D & flexible 3D  & flexible 3D \\
\hline Memory   & 256 MB & 8 GB &  64 GB & 1 TB \\
\hline \# registers (width)  & 64 (32 bit)  & 128 (32 bit)  & 512 (32 bit)  & 512 (64 bit) \\
\hline clock speed   &  8 MHz  & 25 MHz   & 66 MHz  &  200 MHz \\
\hline Total Computing Power of all &   ~1.5 GFlops & ~ 250 GFlops & ~ 2 TFlops & ~ 8-20 TFlops \\
\hline
\hline
\et
%%%%\vspace*{-0.8cm}
\end{table*}

%----- =========================== ------
\begin{figure}[t]
\bc
\includegraphics[width=7.5cm]{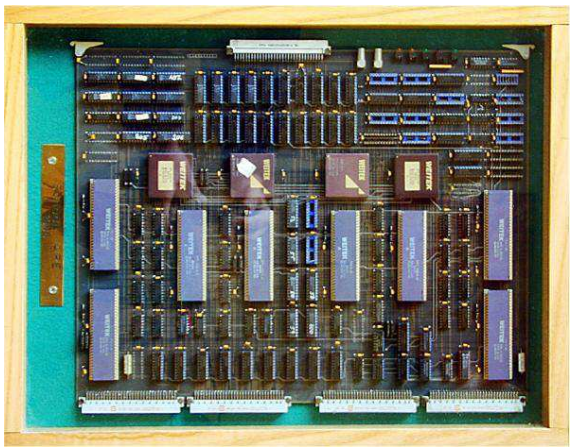}
\ec
\caption{One of the original 1988 \ape{} boards.\label{fig:APE}}
\end{figure}
%----- =========================== ------

The evolution over more than one decade of \ape{} systems is briefly
recollected in Table \ref{tab:apefamily}.

\begin{figure}[t]
\bc
\includegraphics[width=7.5cm]{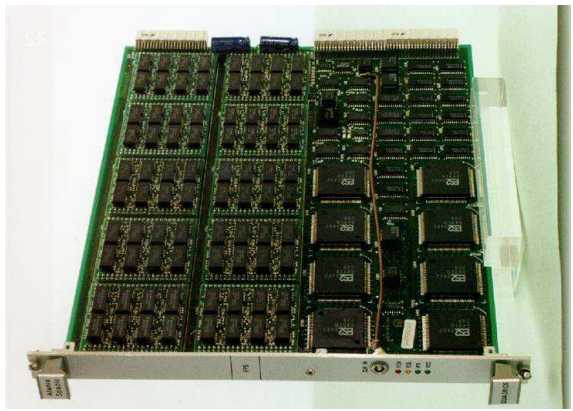}
\ec
\bc
\includegraphics[width=7.5cm]{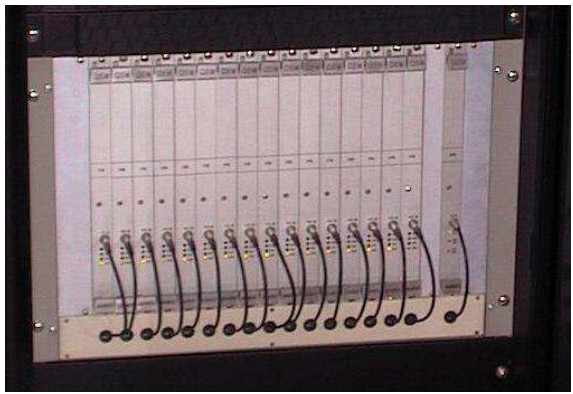}
\ec
\caption{A \acento{} board and a 6.4 GFlops \acento{} crate 
operating at Parma University.\label{fig:APEcento}}
\end{figure}

The first generation of \ape{} computers dates to the mid eighties. In
Fig.~\ref{fig:APE}, a picture of the original \ape{} processor, made out of
off-the-shelf electronic components is shown as a historical remark.
\acento{}, the second
generation of \ape{} supercomputers, had been the leading workhorse of
the European lattice community since the middle of the 1990s.
Several parts of the \acento{} machine are shown in Fig.~\ref{fig:APEcento}.

Commissioning of \amille{}, the third generation of \ape{} systems, started in the year 2000.
These machines make a further 2 TFlops of computing power available to
the LGT community. A description of the \amille{} architecture is given in
a later section.

%----- =========================== ------
%----- =========================== ------
In order to keep up with future and growing requirements, the
development of a new generation of a multi-TFlops computer for LGT,
\anext{}, is in progress.  The main goal \cite{proposal} is the
development and commissioning of a supercomputer with a peak
performance of more than 5 TFlops and a sustained efficiency of
$O(50\%)$ for key lattice gauge theory kernels. Aiming for both large
scale simulations with dynamical fermions and quenched calculations on
very large lattices the architecture should allow for large on-line
data storage (of the order of 1 TByte) as well as input/output channels
which sustain at least $O(0.5)$ MByte per second per GFlops.
Finally, the programming environment should allow smooth
migration from older \ape{} systems, i.e.~support the TAO language, and
introduce for the first time a C language compiler.

\section{\amille{} SYSTEMS}

%----- =========================== 
\begin{figure}[t]
\bc
\includegraphics[width=7.5cm]{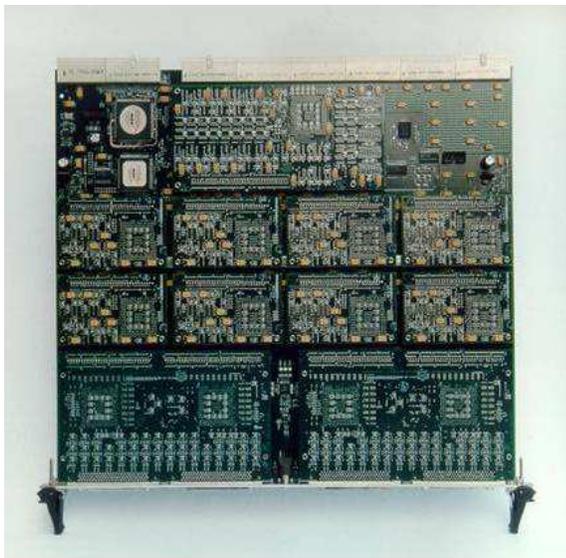}
\ec
\caption{The \amille{} processing board, with its 8 nodes, is the 
smallest possible building block of an \amille{} system.
\label{fig:SchedaAPEmille}}
\end{figure}
%----- =========================== 

%----- =========================== 
\begin{figure}[t]
\bc
\includegraphics[width=7.5cm]{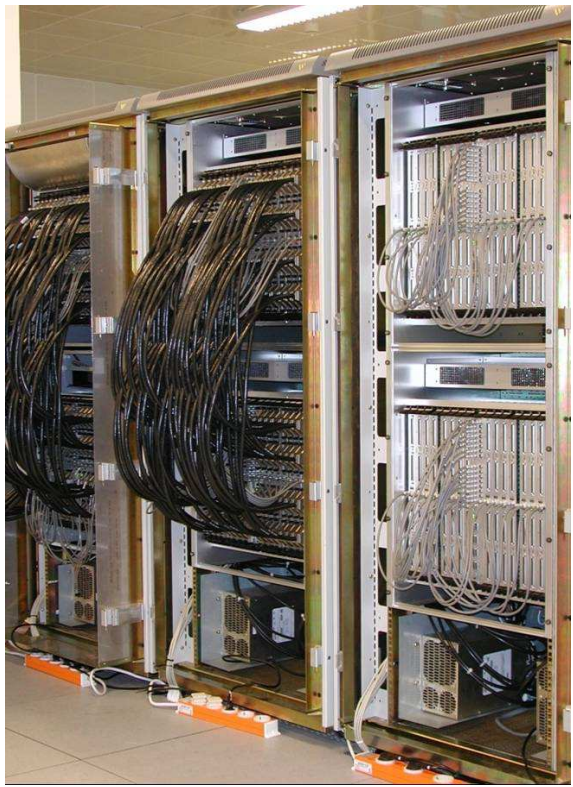}
\ec
\caption{Three \amille{} racks at DESY Zeuthen. Each rack contains
$2 \times 2 \times 2$  computing nodes and has a peak performance 
of 130 GFlops. 
\label{fig:APEmille}}
\end{figure}
%----- =========================== 

\amille{} is a massively parallel computer optimized for simulating
QCD. The architecture is single instruction multiple data (SIMD) 
and all nodes run strictly synchronously
at a moderate clock frequency of 66 MHz.  The communication network
has a three-dimensional topology and offers a bandwidth of 66
MBytes/s/node.  The smallest \amille{} unit 
(see Fig.~\ref{fig:SchedaAPEmille}) is a processing board 
with $2 \times 2 \times 2$ nodes. 
The largest stand-alone systems built until now consist of $4\times 8\times 8$ nodes (see Fig.~\ref{fig:APEmille}).

 Three different integrated circuits (ASICs) have
been custom-developed for \amille{}. Program execution is controlled by
a control processor, which also performs the subset of integer
arithmetics common to the whole SIMD partition. Computations using
local integers and all floating-point operations are done in parallel
by all computing nodes. At each clock cycle, the arithmetic processors
are able to complete the ``normal'' operation $a\times b +c$, where
$a$, $b$ and $c$ are single precision (32 bit) complex operands. This
gives a peak performance of 528 MFlops per node. Each node has 32
MBytes of local memory and a very large register file, holding up to
512 data words. Remote communications between the nodes are
implemented as direct memory access which is controlled and routed by
the communication processors.  The processors are controlled by very
long instruction words (VLIW). This allows efficient scheduling of
the microcode at compile time. Much effort has therefore been put
into the development of software tools for generating efficient code
(see later for more details). Loading of the executables and all other
operating system services are handled via PCs running Linux. One host
PC per four boards is directly attached to the \amille{} backplane. It
uses a PCI bus to communicate with the processing boards. The user
interfaces with the system by logging onto an front-end PC, which is
also running Linux and allows to spawn the program on \amille{} and
to monitor its execution.
Altogether, a large \amille{} installation may be controlled by more than 20 PC's.
These are connected via an Ethernet network.

\amille{} machines are installed at several sites all over Europe, as
detailed in table~\ref{tbl:amille_installations}. They provide a very stable and
reliable computing environment, with typical up-times of the order of
85 \%.
\begin{table}[t]
\caption{\label{tbl:amille_installations}The \amille{} installations.}
\bt{|l|rl|}
\hline
~~Bielefeld ~~&~~ 130 GFlops ~~& ~~(2 crates)    ~~\\
~~Zeuthen   ~~&~~ 520 GFlops ~~& ~~(8 crates)    ~~\\
~~Milan     ~~&~~ 130 GFlops ~~& ~~(2 crates)    ~~\\
~~Bari      ~~&~~  65 GFlops ~~& ~~(1 Crates)    ~~\\
~~Trento    ~~&~~  65 GFlops ~~& ~~(1 Crates)    ~~\\
~~Pisa      ~~&~~ 325 GFlops ~~& ~~(5 Crates)    ~~\\
~~Rome 1    ~~&~~ 520 GFlops ~~& ~~(8 Crates)    ~~\\
~~Rome 2    ~~&~~ 130 GFlops ~~& ~~(2 Crates)    ~~\\
~~Orsay     ~~&~~  16 GFlops ~~& ~~(1/4 crates)  ~~\\
~~Swansea   ~~&~~  65 GFlops ~~& ~~(1 crates)    ~~\\
\hline
\et
%%%%\vspace*{-0.8cm}
\end{table}

%-------------------------------------------------------------------------------

\section{apeNEXT PROCESSOR AND GLOBAL DESIGN}

%Although there are a number of similarities between the architecture
%of \anext{} and former generations of \ape{} supercomputers, there
%were a number of design challenges to be solved in order to meet the
%machine specifications outlined above.

%%%%%%%%%%%%%%%%
\anext{} has been designed with the main goal of having an architecture
as close as possible to its previous generation, as perceived at the
user level, while improving on performance as much
as possible through the use of more advanced technology. 
As a consequence, there are many similarities with previous generation
systems, while a number of new design challenges had to be solved.

For \anext{} all processor
functionalities, including the network devices, are integrated into
one single custom chip running at a clock frequency of 200 MHz. Unlike
former machines, the nodes will run asynchronously, which means that
\anext{} follows the single program multiple data (SPMD) programming
model.

Our performance goal is based on a peak performance of 1.6 GFlops per 
processor in 64-bit double precision, while the communication bandwidth 
between neighboring nodes is 200 MByte/s. We envisage large \anext{} systems
with 2000 processing nodes, delivering a peak performance of 3.2
TFlops. The key design parameters are listed in table~\ref{tbl:anext_parameter}.

\begin{table}[t]
\caption{\label{tbl:anext_parameter}Key \anext{} parameters.}
\bt{|l|l|}
\hline
clock frequency    & 200 MHz \\
peak performance   & 1.6 GFlops \\
memory             & 256-1024 MByte/node \\
memory bandwidth   & 3.2 GByte/sec \\
network bandwidth  & 0.2 GByte/sec/link \\
register file      & 512 registers \\
instruction buffer & 4096 words \\
\hline
\et
\end{table}

Two new key features have been introduced in \anext{}.
First, \anext{} is a SPMD (as opposed to SIMD)
system. Each processing node is a fully independent processor, with a
full-fledged flow-control unit and, of course, a number-crunching
unit. The node has access to its own memory bank, where both program
and data are stored. It executes its own copy of the program at its
own pace. Nodes need to be synchronized only when a data-exchange
operation is performed. This architecture may be labeled as a
distributed-memory parallel computer, in which nodes exchange data
through some sort of ``message-passing'' scheme. Internode
communications are started by the program on the sending node that
initiates a data communication step. This operation is matched by a
corresponding instruction on the destination node, that explicitly
receives the data packet. The latency associated to a ``message'' is
extremely short, of the order of 2 to 3 times the latency associated
to an access to local memory. For this reason, the actual data rate
between nodes is bandwidth-limited (as opposed to latency-limited)
even for short packets, so sequences of short accesses can be freely
programmed without significant performance losses. 
This is an important feature in LGT, where the natural size of
data packet transferred to remote processors is not larger than about
200 bytes.

A second important architectural enhancement lays on the
possibility of routing all read memory accesses (to local or remote
nodes) through a receiving queue, which can be later accessed by the
processor with zero latency. This feature is mainly used to perform
data pre-fetch in critical kernel loops, taking into account
that the address patterns are regular and easily predictable. 
The basic idea here is that
all data items needed to perform iteration $(i+k)$ ($k=1,2, ...$) of the loop are
pre-fetched during iteration $i$ and stored into the queue. When
iteration $(i+k)$ starts,  data will be immediately available to the
processor, effectively hiding almost all latency effects.  Note that some of
the memory accesses will be local, and some will be remote. They are
started in sequence, but they may complete in a different order (a
remote access may take longer than a local one).  However, the queue
mechanism automatically ensures that data are delivered to the
processor in the same order in which they were requested from (remote
or local) memory.

\begin{figure}[t]
\bc
\vspace*{0.1cm}
\includegraphics[width=7.5cm]{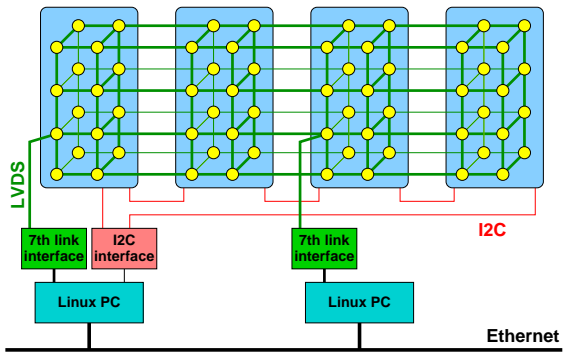}
%%%\vspace*{-1.5cm}
\ec
\caption{A possible \anext{} configuration with 4 boards, 2 external
LVDS-links for I/O, and a chained I2C-link for slow-control.
\label{fig:global-arch}}
%%%\vspace*{-0.5cm}
\end{figure}

\begin{figure}[t]
\bc
\includegraphics[width=7.5cm]{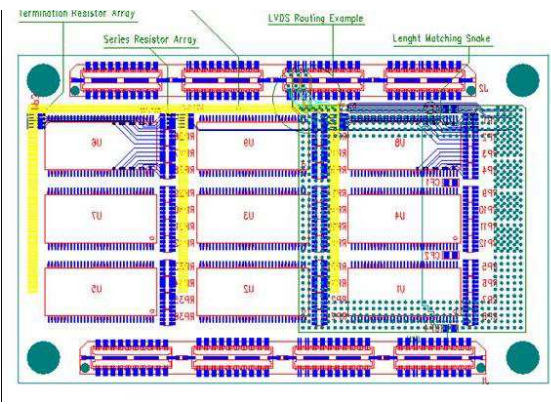}
\ec
\caption{The \anext{} J{\&}T module contains one custom designed VLSI 
J{\&}T chip and nine 8x256 MBytes local DDR-SDRAM chips. 
\label{fig:JTmodule}}
\end{figure}

The complete processing element is contained
in just one custom-designed integrated circuit, called J{\&}T,
connected to a memory bank of 256--1024 MBytes with Double Data Rate
(DDR) Dynamic RAM chips. In turn, the processor and its memory is housed
on a small piggy-back printed circuit board shown in Fig.~\ref{fig:JTmodule}. This assembly
is basically a complete processor, that delivers more than 1.5 GFlops of
processing power in double precision with a power consumption of approximately
7 Watt, that is about 10 times less than current
generation high-end PCs. The compactness and the low power consumption of this basic building
block are key ingredients to build the very compact multi-node system
described later on.

%%%%%%%%%%%%%%

\begin{figure}[t]
\bc
\includegraphics[width=7.5cm]{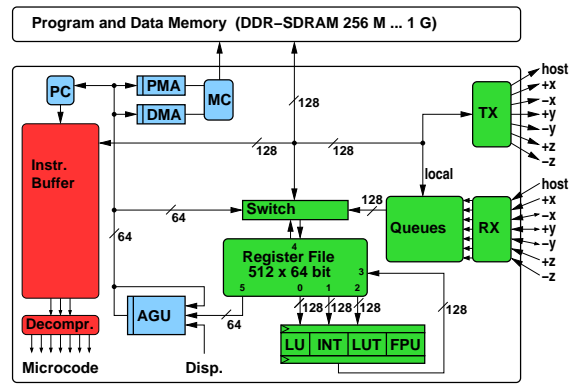}
\ec
\caption{Schematics of the \anext{} J\&T processor.\label{fig:ProcNext}}
\end{figure}

The block diagram of the J\&T chip is shown in
Fig.~\ref{fig:ProcNext}.  It is a 64-bit architecture, optimized for
floating point performance.  
%The basic mathematical operation that can
%be started at each clock cycle is again the so-called \ape{} ``normal''
%operation $a\times b+c$, where $a$, $b$, and $c$ are IEEE double
%precision complex numbers. Each such operation corresponds to eight floating-point
%instructions per clock cycle.

We would like to highlight some selected details of the processor shown in Fig.~\ref{fig:ProcNext}:
%Selected details of the processor shown in
%Fig.~\ref{fig:ProcNext} are described in the following:

\begin{itemize}
\item A large register file of 256 registers each containing a pair
of 64 bit words. All operands for the arithmetic
unit arrive from the register file and all results are
written back here.

\item An arithmetic box which performs floating point as well as
integer operations. The basic operation executed by the floating
point unit (FPU) is the ``normal''
floating-point operation $a\times b+c$. At each clock cycle
one normal operation can be started to provide a maximal 
throughput of eight floating-point operations per clock cycle.
The operands are either complex values or pairs of real values.
All floating-point data is represented in the 64-bit double-precision 
format of the IEEE standard.
The integer unit (INT) operates on pairs of 64-bit integers.
The arithmetic box also contains the logical unit (LU).
Finally, a ``special function'' block (LUT) provides initial approximations
for iterative or series-expansion evaluation of inverses, square-roots,
logarithmic and exponential functions.

\item An address-generation unit (AGU) which computes
addresses for memory access independently and concurrently with the main
arithmetic box. This is an important feature to boost sustained performance.

\item A  memory controller (MC) supporting a memory bank of 256--1024~MBytes
based on standard DDR-SDRAM.
The memory is used to store both data and program
instructions.
%This feature is different from previous
%\ape{} systems where independent program and data memory structures were
%foreseen. In \anext{} this is not possible, since processors are not
%strictly synchronized. 
A consequence of this organization is that
conflicts between data and instruction load-operations
are present.
Two strategies have been employed to
avoid these conflicts. First, the hardware supports compression of
the microcode. The compression rate depends on the level of
optimization, typically values are in  the range of 40--70\%.
Instruction de-compression is performed on-the-fly by dedicated hardware.
Second, an
instruction buffer allows pre-fetching of (compressed) instructions.
Under complete software control, a section of the instruction buffer can 
be used to store performance critical kernels for repeated execution.

\item A flow-control unit that executes programs specified as a
sequence of compiler prepared microcode words, using the VLIW control style.
%Microcode sequences corresponding to the executed program are kept in
%main memory. 
%A program cache, where code belonging to critical kernels can be
%kept in the processor, is available however.
%This element is needed to save precious memory bandwidth for data
%access. Program lines are compressed by the compiler and stored in
%main memory and copied onto the cache in compressed form.

\item  A network interface which contains seven LVDS link interfaces.
Each link is bi-directional allowing send and receive operations to
run concurrently.  Once a communication
request is queued it is executed independently of the rest of the
processor, which is a prerequisite for overlapping network and
floating point operations. Each transmitter (TX) is able to send one
byte per clock cycle, i.e.~the gross bandwidth is 200 MByte per second
per link. Due to protocol overhead the effective network bandwidth
is $\le 180$ MByte per second. The network latency is
$O$(0.1~$\mu$s) and therefore at least one order of magnitude smaller
than for today's commercial high performance network
technologies.
While six links are used
to connect the processing node to its neighbors,
the seventh link is used for input/output operations.

\item A set of fifos that implements the queue mechanism described earlier in the text:
\begin{itemize}
\item A first set of fifos (the TX ones) hold the data words received from the main 
memory (or the register file) until the network is able to send them 
to another node. 
The main role of this system is to decouple the (fast) memory system
from the (slower) network links.

\item A second set of fifos receives data items from the network and 
stores them until the processor wants to load them into the register file.

\item Finally, a third set of fifos is used to keep control information.
These fifos are needed to guarantee the correct order of the data bursts.
\end{itemize}

\item A slow serial interface based on the I2C standard,
used for system initialization, debugging and exception
handling.
\end{itemize}

The overall structure of an \anext{} system is similar to \amille{}. We
have a three-dimensional array of 
processors (see Fig.~\ref{fig:global-arch}).
Periodic boundary conditions are applied.
Although all nodes are connected to their nearest neighbors only, the
hardware allows routing across up to three orthogonal links to
all nodes on a cube, i.e.~connecting nodes at distance
$(\Delta_x,\Delta_y,\Delta_z)$ with $|\Delta_i| \le 1$.

\begin{figure}[t]
\bc
\includegraphics[width=7.5cm]{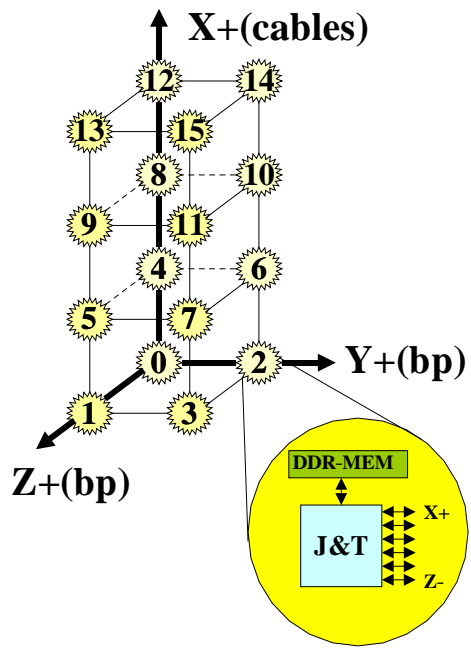}
\ec
\caption{The connection layout of the sixteen J{\&}T nodes in each 
\anext{} processing board. Note that the connection between nodes in 
different processing boards is realized through the back plane for 
communication along the  $Y$ and $Z$ directions and using front-plugged 
cables for communication in the $X$ directions. \label{fig:DiagXYZschem}}
\end{figure}

Clusters of 16 \anext{} processors will be assembled onto just one
printed-circuit board. The processors are arranged in the configuration
of a three-dimensional structure of $4 \times 2 \times 2$ processors
(see fig. \ref{fig:DiagXYZschem}).
A set of $16$ boards is housed within one system crate. 
All communication links between these nodes are established through the crate backplane.
Larger systems are assembled connecting together several crates using external cables.

\begin{figure}[t]
\bc
\includegraphics[width=7.5cm]{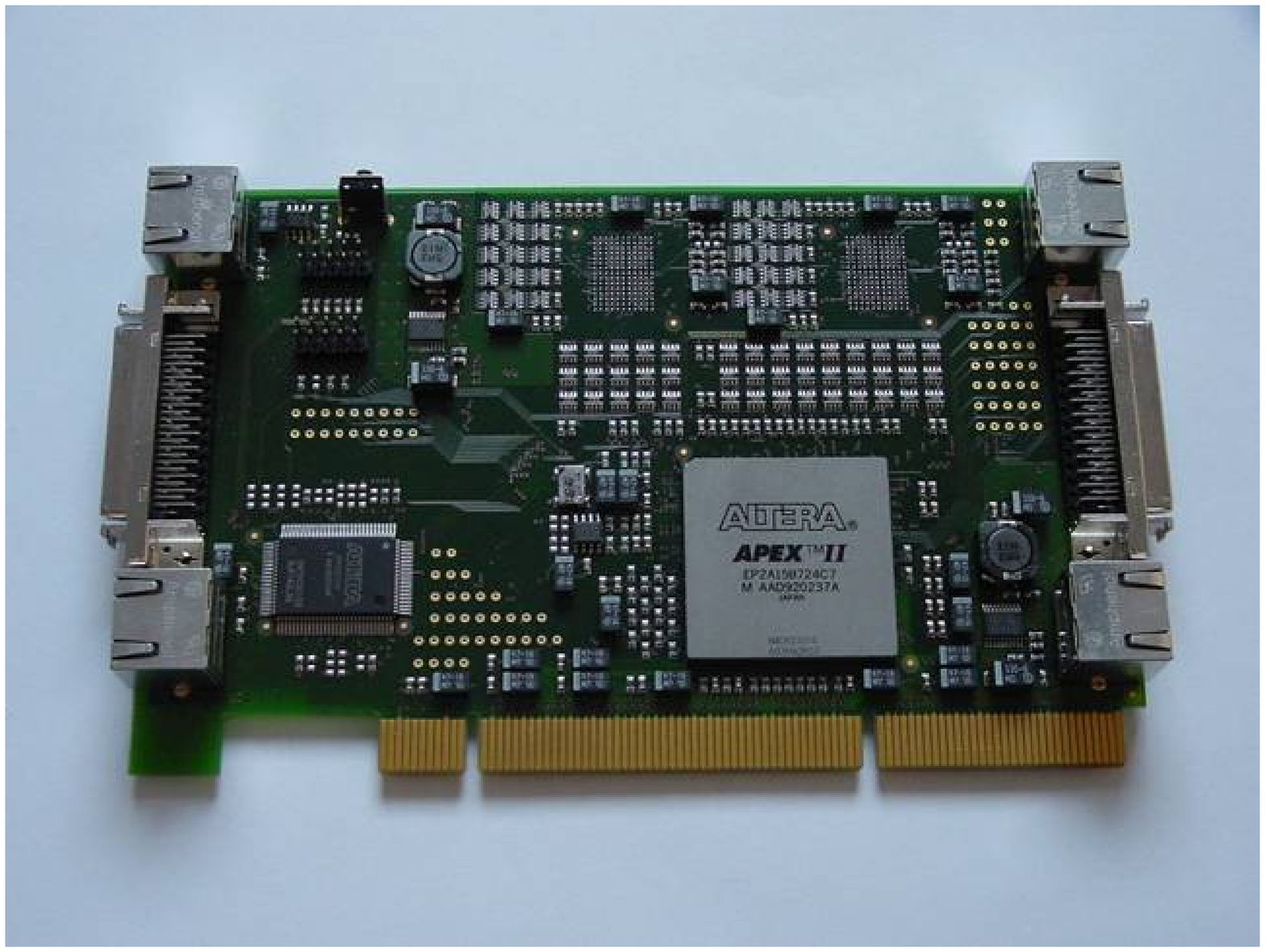}
\includegraphics[width=7.5cm]{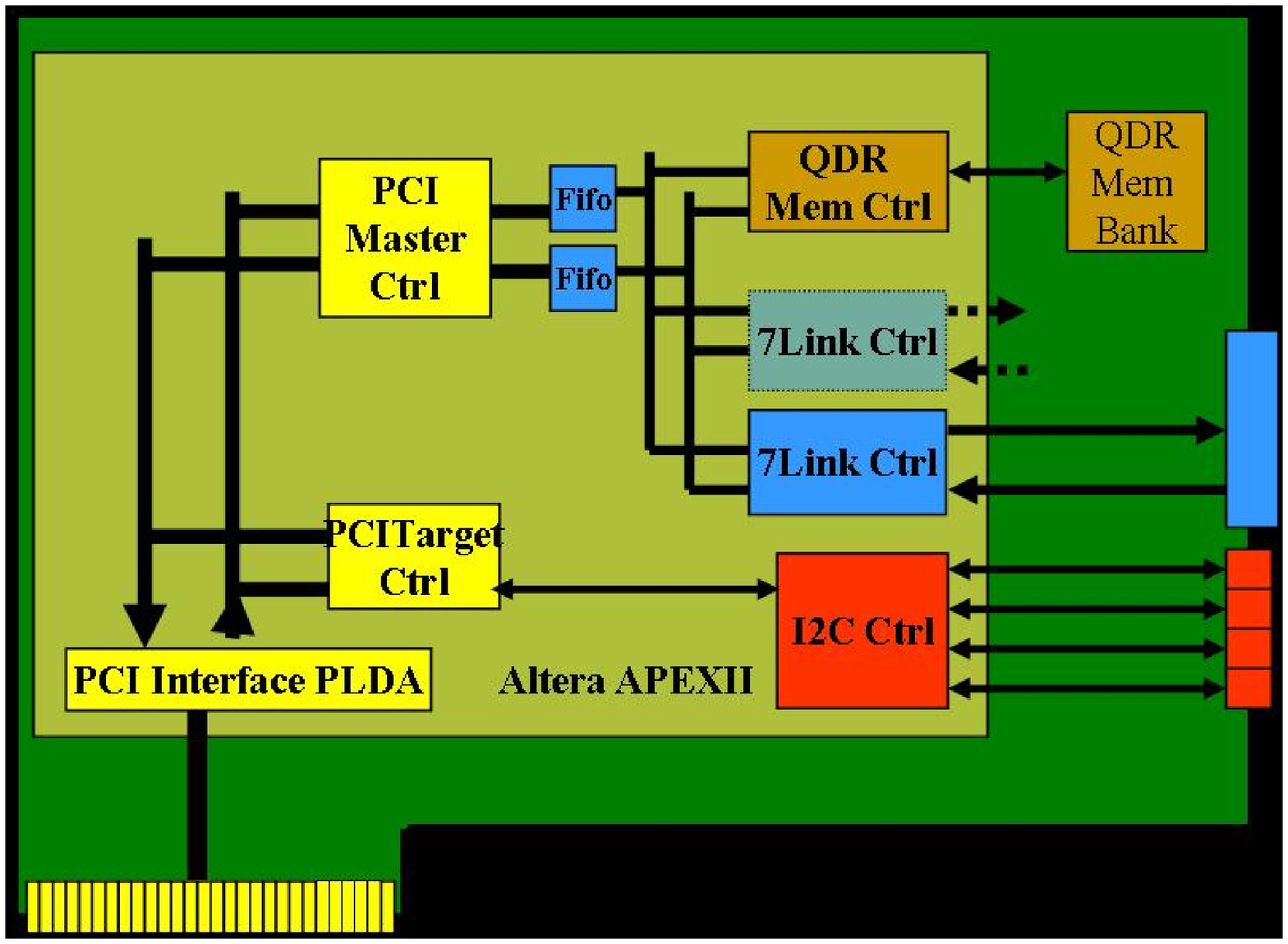}
\ec
\caption{The \anext{} custom made PCI interface. It provide the interface 
between the host Linux PC and the \anext{} system. On board there are 
two 0.2 GByte/sec fast $7^{th}$-link communication channels and
a slow I2C controller.
\label{fig:altera}}
\end{figure}

The system is completed by a number of host PCs that can be tailored to match
user-specific input/output requirements (see fig.~\ref{fig:global-arch}).
The main input/output channel is based on one of the seven data links
available on each processor.
One such link from each processing board
can be connected to a host interface board that follows the PCI specifications
and can be plugged into a standard PC (see Fig.~\ref{fig:altera}).
The actual number of input/output channels can be 
tailored to match user-specific input/output requirements.
From the \anext{} point of view, an input/output operation is simply 
a remote communication with a special remote node. This structure has 
the main advantage of requiring a minimum of programming and operating 
system overhead.
%-------------------------------------------------------------------------------
\section{SOFTWARE AND BENCHMARKS}

\begin{figure}[t]
\bc
\includegraphics[scale=0.4]{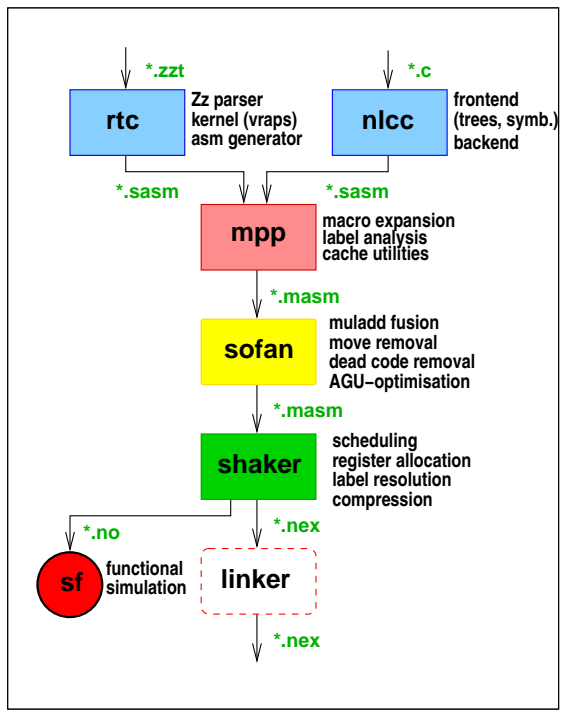}
\ec
\caption{An overview on the compilation procedure.
\label{fig:compiler}}
\end{figure}

We will provide both a TAO and a C compiler for \anext{}. The latter
is based on the freely available {\em lcc} compiler \cite{lcc} and
supports most of the ANSI 89 standard with a few language extensions
required for a parallel machine. Both compilers generate a high-level
assembly. A assembler pre-processor ({\em mpp}) is used to translated
this into a low-level assembly.
For machine specific optimizations
at this assembly level, e.g.~address arithmetics and register move
operations, the software package {\em sofan} is under development.
Finally, the microcode generator ({\em shaker\/}) optimizes
instruction scheduling, which for \ape{} machines is completely done
in software. We finally plan to develop a linker which allows to combine
several microcode files into one executable. An overview on the compilation
chain is shown in Fig.~\ref{fig:compiler}.

For most parts of the compiler software stable prototype versions are
available and were already used to benchmark the \anext{} design.  For
this purpose we considered various typical linear algebra operations,
like the product of two complex vectors. To execute this operation
iteratively needs reading two complex numbers per iteration, which on \anext{}
takes at least two clock cycles. The arithmetic instructions executed
per iteration, however, have a throughput per clock cycle of up to one.
Therefore, this operation is on \anext{} restricted by the memory bandwidth
and the theoretical sustained performance therefore limited to 50\%.
From VHDL simulations that include all machine
details the efficiency was found to be 41\%. Even higher performance
rates can be achieved for operations where the data after being loaded to
the register file is re-used several times. This is, e.g., the
case when multiplying arrays of SU(3) matrices. For this operation we
measured an efficiency of 65\%. 

In QCD simulations
most of the time is spent applying the Dirac operator, e.g.~the
Wilson-Dirac operator $M=1-\kappa H$.  Therefore, we investigated the
multiplication with the so-called hopping term, i.e.~$H \psi$. 
This operation involves remote communications and therefore depends
on the number of processors involved. The maximum number of processors
is limited by the size of the problem, i.e. the lattice volume.
In case of dynamical fermion simulations a multiplication with the
hopping term has to be done much more often than updating the gauge
fields. We therefore kept a local copy of the gauge fields to save
network bandwidth. Considering the worst case where the problem is
distributed over the maximum number of processors, we found the
sustained performance to be 56\%.  This figure is made possible by
extensive use of the pre-fetch features of the processor, which allowed 
to completely overlap floating point operations and network
communications, such that the time when the processor waits for
data becomes almost zero. This eventually indicates an excellent
scaling behavior of the \anext{} architecture.

\section{OUTLOOK AND CONCLUSIONS}

The hardware design of the next generation of \ape{} custom built
computers has been completed. 
Prototype boards and the communication backplane are
already available. 
A prototype \anext{} processor is
expected out of the foundry in late August 2003. 
A larger prototype installation is
planned to be running by end of 2003. There exists a stable
prototype version for all parts of the compiler software. Based on
this software we expect that key lattice gauge
theory operations will be able to run at a sustained performance of
$O(50\%)$ or more. We hope that \anext{} will be a key element of
LGT simulations in the next few years.

\begin{acknowledgments}
We would like to thank W. Errico for his important work in the early
phases of the project, and A.~Agarwal, T.~Giorgino and M.~Lukyanov 
for their contributions.

\end{acknowledgments}


\end{document}